\documentclass[aps, prl, reprint, amsmath, amssymb,superscriptaddress]{revtex4-1}

\usepackage{bm}
\usepackage{bbm}
\usepackage[retainorgcmds]{IEEEtrantools}
\usepackage{graphicx}
\usepackage{mathrsfs}
\usepackage{braket}
\usepackage{amsmath}
\usepackage{amsfonts}
\usepackage{amssymb}
\usepackage{color,xcolor}
\usepackage{times,txfonts}
\usepackage{nicefrac}
\usepackage{ragged2e}
\usepackage{float}
\usepackage{tikz}

\usepackage[colorlinks=true,linkcolor=blue,urlcolor=blue,citecolor=blue,pdfusetitle]{hyperref}

\usepackage{blindtext}
\usepackage[all]{hypcap}

\begin{document}

\title{Counting observables in stochastic excursions}
\date{\today}
\author{Guilherme Fiusa}
\email{gfiusa@ur.rochester.edu}
\affiliation{Department of Physics and Astronomy, University of Rochester, Rochester, New York 14627, USA}
\author{Pedro E. Harunari}
\affiliation{Complex Systems and Statistical Mechanics, Department of Physics and Materials Science, University of Luxembourg, 30 Avenue des Hauts-Fourneaux, L-4362 Esch-sur-Alzette, Luxembourg}
\affiliation{Aix Marseille Université, CNRS, CINAM, Turing Center for Living Systems, 13288 Marseille, France}
\author{Abhaya S. Hegde}
\affiliation{Department of Physics and Astronomy, University of Rochester, Rochester, New York 14627, USA}
\author{Gabriel T. Landi}
\affiliation{Department of Physics and Astronomy, University of Rochester, Rochester, New York 14627, USA}

\begin{abstract}
Understanding fluctuations of observables across stochastic trajectories is essential for various fields of research, from quantum thermal machines to biological motors. 
 {We introduce a framework to analyze the statistics of counting observables in sub-trajectories---dubbed as \emph{stochastic excursions}---of processes out of equilibrium.}
 {Given a partition of the state space into two sets $A$ and $B$, an excursion is defined as the segment of the trajectory that starts with a transition from $A$ to $B$ and ends upon the first return from $B$ to $A$.}
 {Our approach offers analytical expressions for the full distribution of counting observables (such as currents, heat, work, entropy production, and dynamical activity) and the excursion duration, capturing their correlations and finite-time fluctuations.} 
As our main result, we uncover a nontrivial fundamental relation between fluctuations of counting observables at the single-excursion level and the steady state noise obtained from full counting statistics, offering a tool to inspect noise sources.
Interestingly, individual excursions preserve a fluctuation theorem structure and satisfy a thermodynamic uncertainty relation, whose insights on precision-cost trade-offs are explored.
We discuss examples from distinct fields in which the excursion framework naturally addresses relevant questions, and explore in more detail how analyzing excursions yields additional insights into the operation of the three-qubit absorption refrigerator.
\end{abstract}

\maketitle{}

\begin{figure*}[!htbp]
    \centering
    \includegraphics[width=\linewidth]{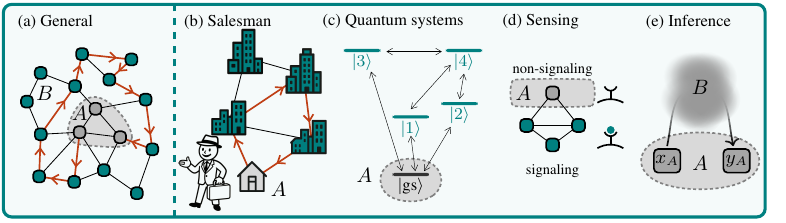}
    \caption{ {\emph{The general setup of stochastic excursions}: (a) A system undergoes stochastic dynamics on a finite set of states that are split into two regions, $A$ (gray) and $B$ (green). 
    An excursion starts with a transition $A\to B$ and ends with the first transition $B\to A$ (red arrows). 
    The states in $A$ do not have to be connected or clustered.
    \emph{Examples of excursions in relevant problems}:
    (b) The salesman performs excursions leaving home and visiting cities. While questions related to time are interesting, a comprehensive characterization of the process requires counting observables such as the amount of money earned or distance traveled per excursion and their interplay with excursion times.
    (c) Quantum thermodynamic models such as heat engines or clocks are often be modeled as a classical Markov jump process. Excursions around the ground state provide information of cyclic heat transfer, work extraction and cost of ticks through counting observables.
    (d) Cellular sensing is often modeled as a stochastic process where some states represent unbound receptors while others represent bound. Subsequent occurrences of unbound receptors mark excursions through bound states. Key quantities such as the environmental concentration of ligands are learned from the properties of such excursions.
    (e) If states in $B$ are hidden, trajectories within it form excursions; if duration and counting observables are empirically available at the end of excursions, their statistics contains information for inferring properties of $B$.
    }}
    \label{fig:drawing}
\end{figure*}

{\bf \emph{Introduction}}---A wide variety of phenomena in nature can be characterized by two alternating phases: an ``inactive'' phase $A$ and an ``active'' phase $B$.
Every once in a while the system becomes activated (e.g., when some energy enters the system), transitioning from $A \to B$. It then spends some random time in $B$ (e.g. performing some task) and eventually returns from $B\to A$, see Fig.~\ref{fig:drawing}(a). 
We term this sequence of events  a \emph{stochastic excursion}.
Problems of this form have been widely studied across various fields, including biology~\cite{bergPhysicsChemoreception1977,  Mora2010, Fancher2020, Harvey2023, Barato2015, Lang2014}, physics~\cite{Prech2024, Hegde2025}, chemistry~\cite{bialekPhysicalLimitsBiochemical2005}, and applied mathematics~\cite{kleinrock1974queueing, Lindley1952, gross2011fundamentals, Bhat2008}. 
The concept of an excursion of trajectories that leave and return to a single reference state has been explored before~\cite{Blumenthal1992}. The framework presented here relies on a similar filtering of trajectories, but is distinguished in three aspects: it applies to Markov jump processes rather than Brownian or Langevin dynamics; it is centered on counting observables, which are defined through the discrete transitions; the final time is a random variable, in contrast to Brownian excursions or bridges where the excursion would end at some fixed later time~\cite{Chung1976, Takacs1991}. Prior works have largely addressed time-integrated functionals and first passage times~\cite{Yao1985, Rubino1989, Kook1993, Takacs1974, Garrahan2017, Bebon2024, Roldan2019}, excursion areas and avalanches~\cite{Majumdar2005, Majumdar2008, Majumdar2015, Mazzolo2017, Zapperi2005, Papanikolaou2011, Baldassarri2021, Meerson2020, Abundo2017, Kearney2007}, and connections with killed and reward renewal processes~\cite{tan2021markovrewardsprocessesimpulse}; but transition-resolved counting observables have been largely overlooked. This is where our framework rests.

As a motivating example, consider a salesman that performs excursions leaving home ($A$) and visiting several cities ($B$) selling goods, see Fig.~\ref{fig:drawing}(b). 
Questions related to time, such as ``how long on average does the salesman take to return home?'' are certainly interesting.
Equally interesting are questions such as 
``how much money did the salesman earn, how many visited cities, what was the total distance traveled?''
Those are all \emph{counting observables}, 
and extending the notion of stochastic excursions to account for them will be the overarching goal of this Letter. 
Even more interesting is the non-trivial interplay between excursion times and counting observables. E.g., if the salesman takes longer to return home, does that mean he is more likely to earn more or less?

The statistics of counting observables  are addressed in the framework of full counting statistics (FCS), a toolkit to evaluate current and fluctuations of observables in classical stochastic processes~\cite{Esposito2007, Esposito2009, Brandes2008, Saito2008,landi2024a}. 
This framework, however, primarily focuses on long-time properties. For instance, in the salesman analogy, FCS effectively quantifies the long-time average income and its fluctuations.
But if one wishes to assess how those earnings fluctuate within each excursion, then an extension of the full counting formalism is required.

While the traveling salesman example is useful to motivate the main ideas, counting observables in excursions are the natural language in many other physical problems. 
For example, in molecular motors of ATP generation, once the system returns to the state with minimal chemical energy, the dynamics is reset and a new energy conversion begins~\cite{Barato2016, Barato2014,Seifert2012, Kumar2010, Kolomeisky2007, Murugan2014}. 
Here, energy transfers per cycle are the physically meaningful observable: fixed-time averages obscure which part of the trajectory bears the thermodynamic cost.
In heat engines, one is often interested in studying cyclic heat transport, where the state of lowest energy is consistently revisited; heat and work currents are counting observables~\cite{Brandner2015, Hegde2025, Pietzonka2018}. A fixed-time approach would inevitably conflate partial cycles, completed cycles, and idle periods.
Many quantum systems, see Fig.~\ref{fig:drawing}(c), such as quantum dots, three-level masers, and absorption refrigerators have been used to experimentally realize these heat engines~\cite{Josefsson2018,Zhang2022,aamirThermallyDrivenQuantum2025, blokQuantumThermodynamicsQuantum2025}.
Recently, a quantum clock was realized using a quantum dot setup, where each tick is precisely one excursion and the entropy produced per tick, a counting observable, is the main object~\cite{Wadhia2025}.
In cell sensing, cells use membrane-bound receptors to sense the concentration of chemicals and perform chemotaxis~\cite{bergPhysicsChemoreception1977, bialekPhysicalLimitsBiochemical2005, Mora2010}. Bounds on the sensing accuracy are directly obtained from the fluctuations in time spent on signaling states~\cite{Harvey2023}, see Fig.~\ref{fig:drawing}(d). The counting observable of entropy directly provides a cost of the process.
Other kinds of problems where counting observables play a role include thermodynamic inference~\cite{rahavFluctuationRelationsCoarsegraining2007, boEntropyProductionStochastic2014, biskerHierarchicalBoundsEntropy2017, yuDissipationLimitedResolutions2024a, espositoStochasticThermodynamicsCoarse2012, ehrichTightestBoundHidden2021, blomMilestoningEstimatorsDissipation2024, liangThermodynamicBoundsSymmetry2024, Tan2021, harunariWhatLearnFew2022, van_der_Meer2022, van_der_Meer2023, harunariUncoveringNonequilibriumUnresolved2024, ertelEstimatorEntropyProduction2024, Maier2024, Deg_nther2024}, where region $B$ is a hidden region. When the system delves into the hidden region, it performs an excursion until it resurfaces at an accessible state, see Fig.~\ref{fig:drawing}(e). Harnessing excursion properties aid the inference of properties such as violation of detailed balance, topology, and bounds on counting observables.

In this Letter we provide a framework to analyze the statistics of counting observables over excursions on Markov jump processes out of equilibrium.
In particular, we obtain a new relation between fluctuations at the level of a single excursion and the steady state noise obtained from FCS~\cite{Esposito2007, Esposito2009, Brandes2008, Saito2008,landi2024a}, unraveling different noise sources: observable fluctuations, fluctuations of excursion times, and their covariance, thus opening an avenue to new insights into the nature of noisy observables.
We also show how individual excursions satisfy a fluctuation theorem (FT) and a thermodynamic uncertainty relation (TUR)~\cite{Jarzynski1997, Jarzynski2004, seifertEntropyProductionStochastic2005a, DeRoeck2007, Crooks1999, Talkner2007a, Talkner2007b, Talkner2008, Hasegawa2019a, Hasegawa2019b, Campisi2009, Campisi2010, Campisi2014, Andrieux2009, Jarzynski2011, Seifert2012, Esposito2009, Harris2007, Horowitz2019, Nakamura2010}.
The goal here is to introduce the framework, discuss overarching ideas related to stochastic excursions,  {and lay out the noise unraveling.}
In an accompanying paper~\cite{supp}, we provide a suite of technical results, showing how our framework provides a way to compute and characterize distributions of any counting observable with simple analytical formulas. We also discuss relevant examples.

{\bf \emph{Formalism}}---We consider a system that occupies a discrete alphabet of states and evolves stochastically according to a Markovian classical master equation. The probability $p_x$ of finding the system in state $x$ obeys 
\begin{equation}\label{M}
    \frac{dp_x}{dt} = \sum_{y\neq x} W_{xy} p_y - \Gamma_x p_x,\qquad \Gamma_x = \sum_{y\neq x} W_{yx},
\end{equation}
where $W_{xy}$ is the transition rate from $y$ to $x$. The framework is easily extended to the case of multiple transitions between two states~\footnote{ {When different mechanisms induce the same change of states, they can be individually modeled by distinct transitions $\ell$ with their own rates $W_{\ell xy}$, rendering the state-space a multigraph, and they can represent e.g. different reservoirs or chemical reactions.}} and the main findings still hold, as explicitly done in~\cite{supp}.
In vector notation, Eq.~\eqref{M} can be written as $  d|p\rangle / dt = \mathbb{W}|p\rangle$, where $|p\rangle$ is a vector with entries $p_x$ and 
\begin{equation}\label{mathbb_W}
    \mathbb{W} = W- \Gamma = \begin{cases}
        W_{xy} & x\neq y \\
        -\Gamma_x & x = y.
    \end{cases}
\end{equation}
Here and throughout $W$ is the matrix with entries $W_{xy}$ for $x\neq y$, and zeros in the diagonal; conversely, $\Gamma$ is the diagonal matrix with $\Gamma_x$ in the diagonals. Assuming irreducibility of the state space, it follows that Eq.~\eqref{M} has a unique steady state $|p^{\rm ss}\rangle$ which is the solution of $\mathbb{W}|p^{\rm ss}\rangle =0$. 
Eq.~\eqref{M} describes the dynamics at the level of the ensemble. 
Conversely, we can also describe the process in a single stochastic trajectory such as $x_1 \to  x_2 \to x_3 \to \cdots$, where the system spends a random amount of time in a given state before jumping on to the next one.
Such a trajectory is comprised of two elements: a set of states $\{x_1, x_2, x_3, \ldots\}$ that the system navigates, and the residence times $\tau_i$, representing how much time the system spent in state $x_{i}$ before jumping to $x_{i+1}$.

Suppose we split the alphabet of states into two arbitrary regions, denoted by $A$ and $B$. 
We then define an \emph{excursion} as a stochastic trajectory that begins whenever the system jumps from a state in $A\to B$ and ends whenever the system first returns from $B \to A$. 
In between, the system spends a random amount of time in $B$, and navigate over a random number of states in it, see Fig.~\ref{fig:drawing}(a). 

{\bf \emph{Counting observables}}---Our main goal in this Letter is to describe the statistics of counting observables within single excursions.
We define a random variable $\hat{N}_{xy}$ that counts how many times the transition $y \to x$ was observed  {in one excursion}. 
A counting observable is then defined as
\begin{equation}\label{counting_observable}
    \hat{Q} = \sum_{x,y} \nu_{xy} \hat{N}_{xy},
\end{equation}
with generic weights $\nu_{xy}$~\cite{landi2024a}.
 The counting observable (and the random variable $\hat{N}_{xy}$) records the first transition $A\to B$, a random number of transitions within $B$, and the last transition $B \to A$. A wide range of relevant quantities are constructed this way.
For example, $\nu_{x y} = 1, \ \forall x,y,$ yields the dynamical activity $\hat{\mathcal{A}}$, which counts the total number of transitions within one excursion. Dynamical activity plays a key role in uncertainty relations~\cite{Di_Terlizzi2018, Nishiyama2024, Vo2022} and has attracted a lot of interest.

Linear counting observables also describe thermodynamic currents, provided that the corresponding weights are anti-symmetric $\nu_{xy} = -\nu_{yx}$~\cite{Wachtel2015}.
Entropy production, a central quantity in nonequilibrium thermodynamics, is the counting observable
$\hat{\Sigma} := -\sum_{x,y} \beta(E_x-E_y) \hat{N}_{xy}$, with weights 
$\nu_{xy} = -\beta (E_x-E_y)$, where $E_{x/y}$ are the energies of states $x/y$ and $\beta$ is the inverse temperature of the reservoir associated with the transition~\footnote{The entropy production associated with transition $y \to x$ is by definition given by $\log{(W_{xy}/W_{yx})}$. Since transition rates generated by thermal reservoirs satisfy local detail balance, it follows that $W_{xy} / W_{yx} = \exp \{ -\beta(E_x - E_y) \},$ where $E_{x/y}$ are the energies of levels $x/y$ and $\beta$ is the inverse temperature of the reservoir associated with it. Therefore, the entropy production of the jump becomes simply $-\beta(E_x - E_y)$, hence the corresponding weights of the counting observable should be $\nu_{xy} = -\beta(E_x - E_y)$.}.
Other fundamental thermodynamic currents, such as heat and work, can be constructed similarly.

{\bf \emph{Excursion statistics}}---Under the split state space,
the transition matrix~\eqref{mathbb_W} acquires the block structure 
\begin{equation}\label{block_W}
    \mathbb{W} = \begin{pmatrix}
\mathbb{W}_{\!\!A} & W_{\!\!AB} \\
W_{\!\!BA} & \mathbb{W}_{\!\!B}
\end{pmatrix},
\end{equation}
with the diagonal blocks
$\mathbb{W}_{\!A} = W_A- \Gamma_A$ and $\mathbb{W}_{\!B} = W_B - \Gamma_B$. 
The diagonal blocks are not stochastic matrices since their columns do not add up to zero.
Next, consider a set of $r$ counting observables $\hat{Q}_\alpha$, defined as in Eq.~\eqref{counting_observable}, 
each with its own set of weights $\nu_{xy}^{\alpha}$. 
Introduce tilted matrices 
\begin{equation}\label{tilted_transition_matrix}
    (\mathbb{W}_{\bm{\xi}})_{xy} =
    \begin{cases}
    W_{xy} e^{-i \sum_\alpha \nu_{xy}^\alpha \xi_\alpha} & x\neq y
    \\
    -\Gamma_x & x = y,
    \end{cases}
\end{equation}
with $r$ counting fields $\xi_\alpha$. 
 {Since excursions are constrained only by the initial and final transitions and the visited states, the duration of an excursion is random variable that we denote by $\hat{T}$.}
For an excursion starting in $x_A$ and ending in $y_A$, the joint probability density that each observable $\hat{Q}_\alpha$ takes on a value $q_\alpha$ \emph{and} total excursion time is $\hat{T} = t$ reads
\begin{equation}\label{Pqt}
    P(\bm{q},t) = C_{x_A \to y_A} \int\limits_{-\infty}^\infty \frac{d\xi_1\ldots d\xi_r}{(2\pi)^r} \langle y_A| W_{AB\bm{\xi}} e^{\mathbb{W}_{\!B\bm{\xi}}t} W_{BA\bm{\xi}} |x_A\rangle e^{i \bm{q}\cdot \bm{\xi}},
\end{equation}
where $\bm{q} = (q_1, \ldots, q_r)$, $\bm{\xi} = (\xi_1, \ldots, \xi_r)$, and 
$C_{x_A \to y_A}^{-1} = \langle y_A|W_{AB} (-\mathbb{W}_B^{-1}) W_{BA}|x_A\rangle$ is a normalization constant, which is proportional to the probability that an excursion follows the path $x_A\to y_A$. For a proof of Eq.~\eqref{Pqt}, see the companion paper~\cite{supp}.
All other calculations follow from this result.
Note that if the span of $\hat{Q}_{\alpha}$ is discrete, then Eq.~\eqref{Pqt} will result in weighted deltas centered around each point over the span of the distribution.
The counting fields $\bm{\xi}$ pick up all events that occur during an excursion. 
We first count the starting jump from $A\to B$ using $W_{BA\bm{\xi}}$. 
Then we count an arbitrary number of jumps within $B$ using $\exp \{ \mathbb{W}_{\!B\bm{\xi}}t \}$. And finally we count the jump from $B\to A$ with $W_{AB\bm{\xi}}$. 
With the tilted transition matrices~\eqref{tilted_transition_matrix}, each transition $y\to x$ picks up the correct factor of $\nu_{xy}^{\alpha}$ for each counting field. The integral over all $\xi_\alpha$ in Eq.~\eqref{Pqt} then takes us from the characteristic function to the actual probability distribution. 

Marginalizing over time $t$ provides the distribution of counting observables irrespective of excursion durations $P(\bm{q})$,
and marginalizing Eq.~\eqref{Pqt} over $\bm{q}$ provides the PDF of excursion durations:
\begin{equation}\label{Pt}
    P\big(\hat{T} = t\big) = C_{x_A\to y_A}\left\langle y_A \left\lvert W_{AB} e^{\mathbb{W}_{\!B} t} W_{BA}\right\rvert x_A\right\rangle,
\end{equation}
a minor generalization of results in first-passage times~\cite{skinner2021}. 
Here we also discuss the physical meaning of the constant $C_{x_A \to y_A}$ that appeared in Eqs.~\eqref{Pqt} and~\eqref{Pt}.
Let $P[x_A \to z_B \vert x_A]$ denote the probability that an excursion starts with a jump $x_A \to z_B$ conditioned on starting in $x_A$. It is given by
\begin{equation}\label{C_PzBxA}
    P[x_A\to z_B \vert x_A] = \frac{\braket{z_B|W_{BA}|x_A} }{\braket{1_B|W_{BA}|x_A}},
\end{equation}
where $\bra{1_B}$ is a vector of length $|B|$ with all entries equal to 1.
The denominator $\mathcal{K}_{BA}^{x_A} \coloneqq \braket{1_B|W_{BA}|x_A}$ ensures normalization. It is precisely the dynamical activity representing the average number of transitions per unit time from $x_A$ to region $B$.

Similarly, let $P[y_A|z_B]$ denote the conditional probability that the excursion ends in $y_A$ given that, when it began, it entered region $B$ through $z_B$ (notice that this is now independent of $x_A$). This is given by 
\begin{equation}\label{C_PyAzB}
    P[y_A|z_B] = \braket{y_A|W_{AB}M|z_B},
\end{equation}
where the factor of $M =-\mathbb{W}_B^{-1}$ accounts for the variable time that the system spends in region $B$ [c.f.~Eq.~\eqref{Pt}]. 
Multiplying Eqs.~\eqref{C_PzBxA} and~\eqref{C_PyAzB} and summing over all states $\ket{z_B}$ yields 
the probability that an excursion ends in $y_A$ given it started in $x_A$:
\begin{equation}
    P[x_A\to y_A \vert x_A] = \frac{\braket{y_A|W_{AB} M W_{BA}|x_A}}{\mathcal{K}_{BA}^{x_A}}.
\end{equation}
Identifying that the quantity in the numerator is precisely $C_{x_A\to y_A}^{-1}$, we arrive at
\begin{equation}\label{C_interpretation}
    C_{x_A\to y_A}^{-1} = P[x_A \to y_A \vert x_A] \mathcal{K}_{BA}^{x_A}.
\end{equation}
Thus, $C_{x_A\to y_A}^{-1}$ relates the probability of the excursion ending in $y_A$ given its start in $x_A$ with the dynamical term $\mathcal{K}_{BA}^{x_A}$, which represents the rate at which excursions originate from $x_A$.
Notice that it only depends on observables estimated from excursions, thereby removing any dependence on the steady state probabilities. 

Simple formulas for analytically calculating the moments of $\hat{Q}_\alpha$ and $\hat{T}$, as well as their correlations, are provided in Ref.~\cite{supp}.
We also provide details on how to evaluate some counting observables of interest such as the dynamical activity. 
Eq.~\eqref{Pqt} considers excursions between two specific states 
$\ket{x_A} \to \ket{y_A}$ in $A$. 
The analysis irrespective of initial and final states amounts to replacing $\ket{x_A} \to \ket{p^{\textrm{ss}}_A}$ and $\bra{y_A} \to \bra{1_A}$  {(also in the normalization constant)}, where $\ket{p^{\textrm{ss}}_A}$ is the part of $\ket{p^{\rm ss}}$ pertaining to region $A$ and $\bra{1_A}$ is a vector with the dimensions of $A$, with all entries equal to 1.
This is an average over initial states and a marginalization over final states.

FTs exist for fixed time trajectories~\cite{Jarzynski1997, Jarzynski2004, seifertEntropyProductionStochastic2005a, DeRoeck2007, Neri2019}, but the excursions preserve such structure, even if working at fluctuating times. Let $P_{x_A\to y_A}(q, \sigma)$ be the probability of a counting observable $q$ with anti-symmetric weights (e.g. a current) and entropy production $\sigma$ in an excursion from $x_A$ to $y_A$, then
\begin{equation}
\label{eq:FT}
    \frac{P_{x_A\to y_A}(q,\sigma)}{P_{y_A\to x_A}(-q,-\sigma)} =  \frac{C_{x_A\to y_A}}{C_{y_A\to x_A}} e^\sigma.
\end{equation}
The factor of $C_{x_A\to y_A} / C_{y_A\to x_A}$ represents the ratio between the probability of an excursion $y_A \to x_A$ and its opposite,
acting as a correcting term of the FT. Following the procedure of Refs.~\cite{Hasegawa2019b, Potts2019}, the FT implies a TUR given by
\begin{equation}
\label{eq:TUR}
    \dfrac{{\rm var}(\hat{Q})_{x_A \to y_A} + {\rm var}(\hat{Q})_{y_A \to x_A}}{\big(E(\hat{Q})_{x_A \to y_A} +E(\hat{Q})_{y_A \to x_A} \big)^2} \geq \dfrac{1}{e^{[E(\hat{\Sigma})_{x_A \to y_A} +E(\hat{\Sigma})_{y_A \to x_A}]/2}-1},
\end{equation}where averages and variances are conditioned over the initial and final states of the constrained trajectories. 
This result shows that the precision-cost trade-off symmetrically constrains the precision of forward and backward excursions: one cannot simultaneously sharpen the fluctuations of a counting observable in both directions without incurring a sufficient entropy cost.
Proofs for Eq.~\eqref{eq:FT} and \eqref{eq:TUR} are detailed in Ref.~\cite{supp}.

{\bf \emph{Connection with steady state FCS}}---Counting observables in FCS are defined as the weighted counting of transitions over a fixed time window $[0,t]$. We shall denote FCS counting observables by $\hat{\mathcal{Q}}(t)$. Explicity, $\hat{\mathcal{Q}}(t)=\sum_{xy}\nu_{xy}\hat{N}_{xy}(t)$ where now $\hat{N}_{xy}(t)$ counts the number of transitions $y \to x$ within the time window. 
In the limit of large $t$, the average current and the diffusion coefficient (noise) are defined as
\begin{equation}\label{FCS_J_D}
    J = \lim\limits_{t\to\infty} \frac{E(\hat{\mathcal{Q}}(t))}{t},
    \qquad 
    D = \lim\limits_{t\to\infty} \frac{{\rm var}(\hat{\mathcal{Q}}(t))}{t}.
\end{equation}
To establish the connection with excursions, we assume that region $A$ has a single state, rendering excursions statistically independent. 
Let $\hat{Q}_n,\hat{T}_n$ denote a counting observable and the excursion time during the $n$-th excursion.
The total \emph{cycle} time of an excursion is $\hat{T}_n^{\rm cyc} = \hat{T}_n + \hat{\tau}_{x,n}$, which includes the residence time  $\hat{\tau}_{x,n}$ that the system spends in $x$ between excursions $n-1$ and $n$. 
Note that $\hat{\tau}_{x,n}$ and $\hat{T}_n$ are independent. 
Finally, let $\hat{\mathcal{N}}(t)$ denote the number of excursions that took place in $[0,t]$. This random variable forms a renewal process~\cite{cox1967renewal}. For sufficiently large $t$, we can then write $\hat{\mathcal{Q}}(t) \simeq \sum_{n=1}^{\hat{\mathcal{N}}(t)} \hat{Q}_{n}$, with the global constraint $\sum_{n=1}^{\hat{\mathcal{N}}(t)} \hat{T}_n^{\rm cyc} \simeq  t$.
The approximation error matters only in the boundary term~\footnote{At time $t$ it may be that $N$ excursions took place, but the system is still halfway through the $(N+1)$-th excursion.}.
It is therefore sub-extensive and can be discarded in the long $t$ limit. 

With this construction in place, we now state the main results (refer to the companion paper~\cite{supp} for the proofs).  
Let $ \mu = E(\hat{T}_n^{\rm cyc}) = E(\hat{T}) + \Gamma_x^{-1}$ and $  \Delta^2 = {\rm var}(\hat{T}_n^{\rm cyc}) = {\rm var}(\hat{T}) +\Gamma_x^{-2}$ denote the mean and variance of the cycle time, with $E(\hat{T})$ and ${\rm var}(\hat{T})$ computed from Eq.~\eqref{Pt}. 
Here we also used the fact that $\hat{\tau}_{x}$ is exponentially distributed with parameter $\Gamma_x$. 
The steady state current in~\eqref{FCS_J_D} is given by 
\begin{equation}\label{fcs-current}
    J = \frac{E(\hat{Q})}{\mu},
\end{equation}
with $E(\hat{Q})$ computed from Eq.~\eqref{Pqt}  {once the joint distribution $P(q,t)$ is marginalized over time}. 
Next, and much less intuitively, the diffusion coefficient in~\eqref{FCS_J_D} is given by 
\begin{equation}\label{fcs-diffusion}
    D = \frac{{\rm var}(\hat{Q})}{\mu} + \frac{\Delta^2}{\mu^3}E(\hat{Q})^2 - \frac{2E(\hat{Q})}{\mu^2}{\rm cov}(\hat{Q},\hat{T}),
\end{equation} where $\textrm{var}(\hat{Q}) = E(\hat{Q}^2)-E(\hat{Q})^2$ and $\textrm{cov}(\hat{Q},\hat{T})=E(\hat{Q}\hat{T})-E(\hat{Q})E(\hat{T})$. We derived closed-form expressions for variances in Ref.~\cite{supp}.
Eq.~\eqref{fcs-diffusion} provides a decomposition of the FCS noise into three mechanistically distinct terms from single-excursion statistics. The first captures the noise arising from fluctuations of the counting observable. The second, the noise due to random excursion durations. And the third, the correlations between an excursion duration and its counting observable. This decomposition has direct physical consequences. It inspects contributions of different quantities to the macroscopic noise, a question that FCS cannot answer, since it delivers $D$ only as a single aggregate number.
As we shall see next in a concrete example, entropy production and dynamical activity can have vastly different noise sources. 

\begin{figure}
    \centering
    \includegraphics[width=1.0\linewidth]{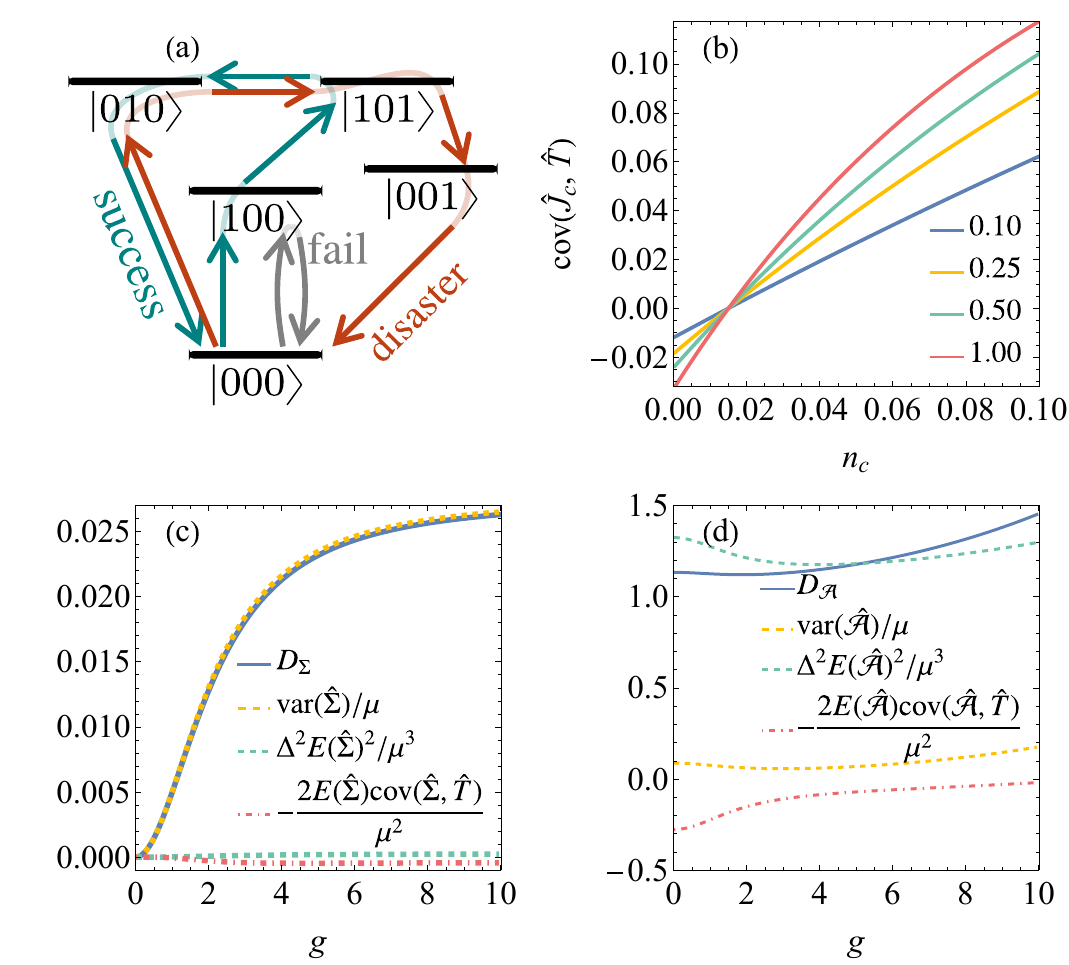}
    \caption{ {(a) Energy diagram of the three qubit absorption refrigerator. Excursions from the ground state $|000\rangle$ can be classified as: success, where the cold current is $+1$; fail, where the cold current is zero; or disaster, there the cold current is $-1$. The different arrows show one example of each type.}
    (b) Covariance between cold current and excursion duration as a function of $n_c$, for different values of $\Gamma_c$ with fixed $g = 10$. 
    (c), (d) Diffusion coefficient (noise) and its decomposition for (c) entropy production, and (d) dynamical activity as a function of $g$, with $n_c = 0.1$ and $\Gamma_c = 0.1$. 
    For all plots, we fixed $\Gamma_h = 50, \ \Gamma_w = 0.5, \ n_h = 0.005,\ n_w = 0.5$.}
    \label{fig:results-refrigerator}
\end{figure}

{\bf \emph{Application: Absorption refrigerator}}--- {Refs.~\cite{aamirThermallyDrivenQuantum2025, blokQuantumThermodynamicsQuantum2025} recently provided experimental demonstration of an absorption refrigerator, which is a paradigmatic model in stochastic and quantum thermodynamics~\cite{Campbell2025}.}
The model consists of three qubits, each coupled to its own heat reservoir, that interact through an effective three-body coupling. 
We consider a simplification with only five energy states, see Fig.~\ref{fig:results-refrigerator}(a), and a discussion of the full model can be found in Ref.~\cite{supp}. 
The five-level model consists of the states $|000\rangle, \ |001\rangle, \ |010\rangle, \ |100\rangle, \ |101\rangle$, where $0$ (resp. $1$) represents that the corresponding qubit is in the ground state (resp. is excited). This ordering of the states is
the convention we use for the transition matrix and the generic
weights, e.g. $W_{41}$ is the rate of the transition $|000\rangle \to |100\rangle$. Region $A$ is taken to be spanned only by the ground state $|000\rangle$.
The quantum dynamics is very well approximated~\cite{supp} by the classical master equation~\eqref{M} in the parameter regime of Ref.~\cite{aamirThermallyDrivenQuantum2025}. 

The transition among states come from two sources, the reservoirs whose transitions rates are described by $\Gamma_\alpha n_\alpha$ for injections and $\Gamma_\alpha (n_\alpha +1)$ for extractions, and the three body interaction $g'$ that couples the states $|010\rangle \leftrightarrow |101\rangle$. 
Reservoirs are taken to be bosonic, so $n_\alpha = \left[\exp({\omega_\alpha}/T_\alpha) - 1\right]^{-1}$ denotes the occupation number and $\Gamma_\alpha$ is the coupling between the qubit and the reservoir.  {$\alpha$ labels the qubit (cold, hot, work).}
The transition matrix reads
\[
W =
\begin{pmatrix}
0 & \Gamma_w (n_w + 1) & \Gamma_h (n_h + 1) & \Gamma_c (n_c + 1) & 0 \\
\Gamma_w n_w & 0 & 0 & 0 & \Gamma_c (n_c + 1) \\
\Gamma_h n_h & 0 & 0 & 0 & g' \\
\Gamma_c n_c & 0 & 0 & 0 & \Gamma_w (n_w + 1) \\
0 & \Gamma_c n_c & g' & \Gamma_w n_w & 0
\end{pmatrix}
,\]
where the classical effective three-body interaction strength is $g'=4g^2/\big(\Gamma_c(n_c+1)+\Gamma_h(n_h+1)+\Gamma_w(n_w+1)\big)$
where $g$ is the quantum three-body Hamiltonian coupling. This result follows from perturbation theory~\cite{supp, Prech2023}. 

Although there are many interesting counting observables to be explored in this model, we restrict ourselves to the cold current, entropy production, and dynamical activity.
The cold current $\hat{J}_c$ is obtained by setting weights $\nu_{xy} = \pm 1$ in Eq.~\eqref{counting_observable} if the transition $x \to y$ removes (resp. adds) an excitation from (resp. into) the cold reservoir. 
 {In excursions over the ground state, there are only three possibilities for the cold current: a ``success'', where cooling took place and the cold current is $\hat{J}_c = 1$; a ``disaster'', where heating took place the cold current is $\hat{J}_c = -1$, and a ``fail'', where $\hat{J}_c = 0$ and there is no cooling nor heating. We sketch excursions of each type in the trajectories drawn in Fig.~\ref{fig:results-refrigerator}(a).}
Excursions can fully characterize the probability distribution of trajectories having zero cold current, in contrast with the steady state analysis~\cite{supp}. 
In Fig.~\ref{fig:results-refrigerator}(b) we show the covariance between the cold current $\hat{J}_c$ and the excursion duration $\hat{T}$ as a function of the cold bath occupation $n_c$. Interestingly, longer excursions tend to have larger $\hat{J}_c$ in the cooling window, which is not necessarily the case when $E(\hat{J}_c) >0$, and a vanishing covariance pinpoints the value of $n_c$ where the cooling window begins~\cite{Mitchison2019}.

In Figs.~\ref{fig:results-refrigerator}(c) and (d), we characterize the noise [Eq.~\eqref{fcs-diffusion}] in terms of its components for the entropy production $\hat{\Sigma}$ and the dynamical activity $\hat{\mathcal{A}}$. In the former, the fluctuations are strongly dominated by the $\text{var}(\hat{\Sigma})$ term; whereas in the latter there is a clear competition between different contributions, with $E(\hat{\mathcal{A}})^2$ being the most relevant term.
This result sheds light on the very nature of the noise. $\hat{\Sigma}$ is an observable that can have positive and negative values related by fluctuation theorems, while $\hat{\mathcal{A}}$ is strictly positive.
The precision of currents is fundamentally constrained by both~\cite{Horowitz2019, Gingrich2016, Vo2022, Timpanaro2019, Hasegawa2019a, Hasegawa2019b, Di_Terlizzi2018}, and the fact that these observables have largely distinct noise sources indicates a subtle and intricate interplay in the thermodynamics of precision.

{\bf \emph{Conclusions}}---We have proposed a framework to characterize the statistics of counting observables in sub-trajectories of Markov jump processes, dubbed stochastic excursions. 
The need for the excursion constraint over trajectories is not merely formal. In many systems of physical interest, such as molecular motors, thermal machines, cellular sensing, and quantum clocks, the excursion is the natural unit of dynamics. The question of what happens within each unit reveals relevant aspects beyond a long-time average analysis.

The decomposition of the FCS noise [Eq.~\eqref{fcs-diffusion}] for $|A| = 1$ reveals a fundamental connection between steady state noise and fluctuations at the single excursion level. Each term is mechanistically distinct, and the third one provides genuine insights into the operation and design of cyclic Markov jump processes. 
Moreover, the framework yields finite-time results (for any $|A|$) that FCS cannot provide: a closed-form joint distribution of counting observables and excursion duration [Eq.~\eqref{Pqt}], a FT [Eq.~\eqref{eq:FT}] that holds without asymptotic approximations, and a TUR [Eq.~\eqref{eq:TUR}] symmetrically relating the precision of counting observables with the entropy produced per excursion. This last result reveals a joint precision-cost trade-off between forward and backward excursions, and reveals that a highly irreversible forward excursion can in principle ``subsidize'' precision in the backward one since an asymmetric irreversibility still constrains the precision. Dissipation governs the precision of the full unit cycle, regardless of the relative probabilities of forward/backward excursions.

The extension of excursions to quantum dynamics is a promising avenue: photon emissions, heat currents, and work exchanges, are counting observables central to quantum thermodynamics, and a self-contained operational cycle is meaningful from circuit QED to quantum-dot heat engines.
A second direction concerns inference in partially observed systems, where region $B$ is hidden and only region $A$ is accessible. The statistics of counting observables and durations measured at the boundary of $B$ carry substantial information about the thermodynamics and topology of the hidden region.

\begin{acknowledgments}
{\bf \emph{Acknowledgments}}---GF and PH acknowledge fruitful discussions with Tan Van Vu, Jann van der Meer, Jiheng Duan, and Pedro Portugal. This research is primarily supported by the U.S. Department of Energy (DOE), Office of Science, Basic Energy Sciences (BES) under Award No. DE-SC0025516. PH was supported by the project INTER/FNRS/20/15074473 funded by F.R.S.-FNRS (Belgium) and FNR (Luxembourg), and by the European Union (ERC-SuperStoc-101117322).
\end{acknowledgments}

\bibliographystyle{ieeetr}
\bibliography{letter}

\end{document}